# The Divergence Between Industrial Infrastructure and Research Output among the GCC Member States

Osman Gulseven*, Abdulrahman Elmi, Odai Bataineh

*In this article, we provide a comparative analysis of the industry, communication, and research infrastructure among the GCC member states as measured by the United Nations sustainable development goal 9. SDG 9 provides a clear framework for measuring the performance of nations in achieving sustainable industrialization. Three pillars of this goal are defined as quality logistics and efficient transportation, availability of mobile-cellular network with high-speed internet access, and quality research output. Based on the data from both the United Nations' SDG database and the Bertelsmann Stiftung SDG-index, our results suggest that while most of the sub-goals in SDG 9 are achieved, significant challenges remain ahead. Notably, the research output of the GCC member states is not in par with that of the developed world. We suggest the GCC decision-makers initiate national and supranational research schemes in order to boost research and development in the region.*

**Introduction**

The Millennium Declaration signed in 2000, stipulated eight Millennium Development Goals (MDGs) to be achieved by 2015 (WHO, 2000). These goals united the world leaders in combating poverty, hunger, and environmental degradation. Research suggests that significant success was achieved in providing social welfare and quality healthcare facilities (WHO, 2000; Liu *et al.*, 2016; Gómez-Olivé and Thorogood, 2018). However, major issues remained intact, particularly in terms of sustainable development, which still remains as a major challenge (Liew, Adhitya and Srinivasan, 2014).

A primary challenge in measuring sustainable development is to provide indicator-based targets where businesses, civic, and government bodies can work on together (Scheyvens, Banks and Hughes, 2016; Allen *et al.*, 2017). The new UN agenda received a considerable demand for such issues. Following these requests, the UN nations held a meeting in 2015, declaring the 17 Sustainable Development Goals (SDGs) to be achieved by 2030 (Brende and Høie, 2015). Known as the UN Resolution 700/1, these SDGs are intended to provide a sustainable future for generations to come. While many of these targets were expected to be accomplished by 2030, some of the targets were ambitiously set to be achieved as early as 2020. There was some ambiguity in how to measure the achievements towards attaining those SDGs. Therefore, an addendum was later amended to the original declaration where, each SDG was defined in several sub-goals, and each sub-goal had several indicators. Finally, a total of 169 targets to be measured by 232 unique indicators are defined in the updated SDG declaration. The 17 SDGs are defined as *no poverty* (#1), *zero hunger* (#2), *good health and wellbeing* (#3), *quality education* (#4), *gender equality* (#5), *clean water and sanitation* (#6), *affordable and green energy* (#7), *decent work and economic growth* (#8), *industry, innovation, and infrastructure* (#9), *reducing inequality* (#10), *sustainable cities and communities* (#11), *responsible consumption and production* (#12), *climate action* (#13), *life below water* (#14), *life on land* (#15), *peace, justice, and strong institutions* (#16), *partnerships for the goals* (#17).

**Dr. Osman Gulseven***, Associate Professor, Skyline University College, UAE, osman.gulseven@skylineuniversity.ac.ae.

**Abdulrahman Elmi , MBA** Student, Skyline University College, UAE
**abdulrahman.elmi@gmail.com**

**Odai Bataineh**, MBA Student, Skyline University College, UAE
**odaibataineh91@gmail.com**

One internationally known ranking system is introduced by the German foundation Bertelsmann-Stiftung (Lafortune *et al.*, 2019). Known as the BS-SDG index, their database compiles data on several indicators for each sub-goal, making it fair and a simple process to compare national achievements within UN members. The measurement of metrics and methodology behind this ranking system has its drawbacks, yet the BS-SDG index is one of the very few metrics that offer a ranking of countries (Miola and Schiltz, 2019). According to BS-SDG rankings, the Nordic-European countries such as Denmark, Sweden, and Finland are ranked as the most sustainable, whereas poor sub-Saharan African nations are at the bottom of the sustainability list. While there is a strong positive correlation between income per capita and sustainability achievements, this is not true for most of the Arab countries – particularly member states of the Gulf Cooperation Council. GCC member states (Bahrain, Kuwait, Qatar, Oman, Saudi Arabia, and UAE) have distinctively low sustainability scored compared to other high-income countries according to the BS-SDG index ranking.

The 2019 ranking puts the United Arab Emirates (UAE) as the top-performing GCC member with a score of 69.7 and a global rank of 65 out of 162 countries. Bahrain has a score of 68.7 and is ranked 76[th] globally. Oman and Qatar have scores of 67.9 and 66.3 and are ranked 83[rd] and 91[st] globally. Saudi Arabia has a score of 64.8 and ranked 98[th], whereas Kuwait is the worst performer among GCC countries with a score of 63.5, and it is ranked 106[th]. Only two of the 17 SDGs are achieved by GCC countries, namely *no poverty* (SDG 1) and *reduced inequalities* (SDG 10). UAE has also achieved SDG 17, *partnerships for the goals*.

One of the main reasons for the low sustainability profile of GCC member countries is the composition of their economic activities. The Gulf region contains the largest proven oil reserves in the world, and as such, these countries are among the top oil producers in the world. The wealth attained by oil production has caused a consumption spree among the newly found rich and middle class. The consumer and environmental awareness in this part of the world is not in par with that of the European countries. This explains the reason for deficient scores in *Responsible consumption and production* (SDG 12), *Climate Action* (SDG 13), *Life below water* (SDG 14), *Life on land* (SDG 15).

However, GCC countries also received low scores in the *industry, innovation, and infrastructure* dimensions (SDG 9). This is quite surprising as these countries have strong industrial ambitions and are known to host business-friendly free zones with no income taxes. In this article, we are analyzing the performance of GCC member states in terms of achieving the sustainable development goal SDG 9. We have chosen to analyze SDG 9 since innovation and industrial infrastructure are leading indicators for forecasting future economic performance. Our research question is whether the economic leap observed in the GCC countries is supported by their achievements in innovation and industrial infrastructure.

**Literature Review on SDG 9**

Sustainable development goals became pillars of achievements around the globe. Each goal is unique, but achievement towards one goal is also related to achievements towards other goals. This relationship can be in either positive or negative ways (Allen, Metternicht and Wiedmann, 2018; Amin-Salem *et al.*, 2018). For example, improvements in economic conditions can simultaneously reduce the inequality within society while combatting hunger and poverty. Digitization and improved technological infrastructure can help to achieve both SDG 3 and SDG 9 (Novillo-Ortiz, De Fátima Marin and Saigí-Rubió, 2018). However, besides synergies, trade-offs are possible. If the improvement in economic conditions happens due to natural resource extraction at excessive rates, it might also deteriorate natural life on land as well as below water. Fighting against climate change is an essential step, but it might mean giving up on carbon-based industrial output for several countries (Campagnolo and Davide, 2019). More research supported with big-data needs to be conducted for a fair assessment of the countries in terms of their achievements towards sustainability (Fullman *et al.*, 2017; Requejo-Castro, Giné-Garriga and Pérez-Foguet, 2020).

While significant research is done in the bioeconomic aspects of the SDGs, very few research is committed to analyzing the role of innovation and infrastructure in achieving sustainable development. One of the reasons is because SDG 9 is somewhat seen as an impediment to achieving success in other SDGs – particularly by bioeconomists. Vinueasa (2020) suggests that innovations such as artificial intelligence can inhibit 58 of the UN SDG targets. Ermgassen et al. (2019) emphasize that the global infrastructure boom shall not turn into a zero-sum game where achievement in SDG means deteriorating life below water and on land. Another research on the correlation of SDGs suggests that there might be significant trade-offs between several SDGs in Europe. According to Ronzon and Sanjuan (2020), there is a 41% negative correlation between SDG 9 and SDG 14 (life below water). SDG 9 is also negatively related to SDG 2, *zero hunger* (37%), and SDG 12, *responsible consumption and production* (37%).

However, synergies are also observed between SDG 9 and other SDGs. A recent report by PricewaterhouseCoopers (2019) suggests that the introduction of SDGs can bring significant challenges as well as opportunities for the global business. According to PwC, the global businesses has the highest potential contribution to SDG 9 by building resilient infrastructure and promoting inclusive, sustainable, and innovative industrialization activities. The same report suggests that the mobile communication industry is driving force behind achieving a connected world, which has universal coverage. Apfelbaum (2018) suggests that the mobile industry is accelerating the delivery of SDGs globally. This unprecedented connection between human society has driven new business models, particularly in areas such as mobile banking and social media. The transportation sector is also a significant contributor to global growth by connecting the economies, and reducing the trade costs while promoting regional integration. This is also emphasized in the OECD report (2018), which emphasizes the role of transport infrastructures such as long-distance roads and railways, international ports, and airports. These are needed to move products, and people around, thereby boosting the trade among countries.

Montes (2017) suggests that SDG 9 is a re-introduction of industrialization into the sustainable development agendas of developing countries, which differ in terms of their population, per capita income, and economic structure. For developed countries, SDG 9 emphasizes clean production systems where carbon emissions affect their achievement scores negatively. The author also claims that the pro-active industry-government partnership is essential to achieve this goal while considering aspects of inequality and sustainability. Several authors recommend similar private-public collaborations to achieve sustainable economic development (Gulseven and Mostert, 2017; Gulseven *et al.*, 2019; Hossen, 2020).

There are several examples where re-thinking industrial growth and proving innovative solutions can boost several SDGs simultaneously. According to Karnama et al. (2019), high-performance computing centers can be redesigned to provide the heat needed for organic food production in greenhouses. The authors point out that by creating this synergy between computing centers and greenhouses, it is possible to progress towards at least 5 of the 17 SDGs, including SDG 9.

The increased use of mobile apps in healthcare monitoring and services has boosted productivity in the healthcare industry. Thus, having a better mobile coverage not only improves the industry infrastructure (SDG 9) but can also boost the quality of health care (SDG 3) in a country (Makri, 2019). Engineering applications – particularly those aimed at improving the life cycle of consumer products can improve most of the SDG indicators (Hauschild, Laurent and Molin, 2018). Research in bioactive compounds can also positively impact both SDG 3 and SDG 9 (Laraia, Robke and Waldmann, 2018).

All in all, there is a consensus view that suggests inclusive and sustainable industrial development is the primary source of income generation, which is needed for a sustained increase in community living standards. As such, sustainable industrialization and innovation indicators are designed to measure the level of transportation infrastructure, connectivity among society, and research & development activities. The following table lists the UN defined sub-goals and indicators for SDG 9.

While some non-academic work has been done on sustainability initiatives in the GCC, there is a shortage of academic work in measuring the performance of these countries in achieving SDGs. Our research adds to the literature by analyzing the performance of GCC states towards achieving sustainable development goal #9, *industry, innovation, and infrastructure*, which is among the low performing SDGs in the region.

**Data**

We retrieve data from the UN Sustainability database. The UN provides a comprehensive database to measure country-wide indicators in achieving SDGs. The SDG handbook also offers a detailed overview of the available data. The UN statistical handbook is produced according to the availability of data in National Statistical Systems. The handbook is still at the development stage, with new data indicators added to the list of SDG indicators database whenever they are available. However, many indicators are defined without paying attention to their availability. Therefore, some of the data were missing. As of December 2019, the currently available information about the indicators is visualized and discussed below. All visualizations are conducted using Tableau software (v2019.1).

**Table 1.** UN Defined Sub Goals and Indicators for SDG 9 Industry, Innovation, and Infrastructure

| Sub-Goal Summary | Indicators As Defined by the Original Declaration | To be achieved by |
|---|---|---|
| 9.1 Develop sustainable, resilient and inclusive infrastructures | 9.1.1 Road access for rural populations<br>9.1.2 Passenger and freight volumes | 2021 |
| 9.2 Promote sustainable industrialization | 9.2.1 Manufacturing value<br>9.2.2 Manufacturing employment | 2021 |
| 9.3 Access to financial services and markets | 9.3.1 Value of small-scale industry<br>9.3.2 Small-scale industries with affordable credit | 2030 |
| 9.4 Resource Efficiency | 9.4.1 CO2 emissions per unit value-added | 2030 |
| 9.5 Scientific Research | 9.5.1 R&D spending<br>9.5.2 Researchers per million inhabitants | 2030 |
| 9.A Aid for sustainable development in developing countries | 9.A.1 Development assistance for infrastructure | 2030 |
| 9.B Support domestic development and industrial diversification | 9.B.1 Medium and high-tech industry | 2030 |
| 9.C Universal access to information and communications technology | 9.C.1 Development assistance for infrastructure | 2021 |

### Target 9.1: Develop sustainable, resilient, and inclusive infrastructures.

*SDG Indicator 9.1.1: Road access for rural populations*

SDG reports do not provide any information regarding this indicator. In developed countries, this is most likely to be around 100%, but there could be a wide variety in developing countries. A better indicator could have been the length of asphalt roads per capita.

*SDG Indicator 9.1.2: Passenger and freight volumes*

While there are several different ways to move passengers and freight, the most robust indicators are about air transport. Thanks to strict rules and statistical data availability, the data on air freights and air passengers are maintained well. The UN defines air freight as the volume of freight bags carried on each flight stage (operation of an aircraft from takeoff to its next landing), measured in metric tons' times kilometers traveled.

As can be seen in Figure 1, the UAE is leading the GCC in terms of freight by air transport. In 2018, this indicator for UAE was 16616 million ton-km. Qatar also leaped in the last decade. In 2018, the freight by air transport indicator was 12667 million ton-km for Qatar. Both of these countries have major globally known airlines supported by large airport bases. Saudi Arabia had only 1085 million ton-km freight, although it is the most populated country in the GCC.

Air passengers carried per year refers to both domestic and international aircraft passengers of air carriers registered in the country. This indicator denotes the number of passengers carried by airliners registered in a given country and is independent of the nationality of individual passengers. GCC countries host significant amounts of expatriate workers. There are also major airlines making a stopover at the major air hubs. Consequently, air traffic in this region is among the highest in the world, as can be seen in Figure 2.

The number of air transport passengers carried has also increased substantially in the last decade. In the UAE alone, 96 million passengers were carried in 2018. This number is 39 million for Saudi Arabia, and 29 million for Qatar. Oman also supported significant air passengers amounting to 10 million people in 2018. The data also shows that Bahrain and Kuwait are not well-known in the airport traveling business.

### Target 9.2: Promote inclusive and sustainable industrialization.

*SDG Indicator 9.2.1: Manufacturing value*

UN defines the manufacturing value-added as the net output of a sector after adding up all outputs and subtracting intermediate inputs in manufacturing production. This indicator is measured as a percentage of the total GDP within the economy.

$$\text{Manufacturing value} = \frac{\text{Manufacturing GDP}}{\text{Total GDP}}$$

Figure 3 shows the progress towards this target, which is measured by measuring the percentage of manufacturing value-added in GDP and per capita. There is a wide range in this indicator among the GCC countries. The UN provided by World Bank development indicators suggests that Bahrain is the most industrialized nation in GCC with a manufacturing GDP ratio of 18%. Bahrain is a small state but runs a mega-sized petroleum refinery, which might be the reason for this high ratio. However, Kuwait also several mage-capacity refineries, but it is the least industrialized with a manufacturing GDP ratio of 7%-8%. Those numbers are highly dependent on how manufacturing is defined. It is also possible that the growth in other sectors of GDP has reduced the role of manufacturing in these economies.

*SDG Indicator 9.2.2: Manufacturing employment*

UN defines manufacturing employment as all persons of working age who, during a specified brief period, were in paid employment (whether at work or with a job but not at work) or in self-employment in a sector categorized as manufacturing. This indicator is measured as manufacturing GDP to total employment.

$$\text{Manufacturing employment} = \frac{\text{Manufacturing employment}}{\text{Total employment}}$$

Figure 4 shows the progress of GCC countries as measured by the proportion of employment in manufacturing out of total employment. The data suggests Qatar has made a leap in the last two decades, and this indicator currently stands at 54%. Oman also made a significant leap with a manufacturing ratio of 36%, followed by Bahrain, which has a manufacturing employment share of 35%. The manufacturing strategy of GCC countries should focus on cross-cutting competencies, vocational training in specific manufacturing sectors, SME linkage programs, and foreign direct investment. We believe that this indicator can not be used to measure the infrastructure in an economy. Investment nowadays means more than old-type manufacturing equipment. Human capital investment is a better indicator.

**Figure 1**. Air Transport, Freight (million ton-km)

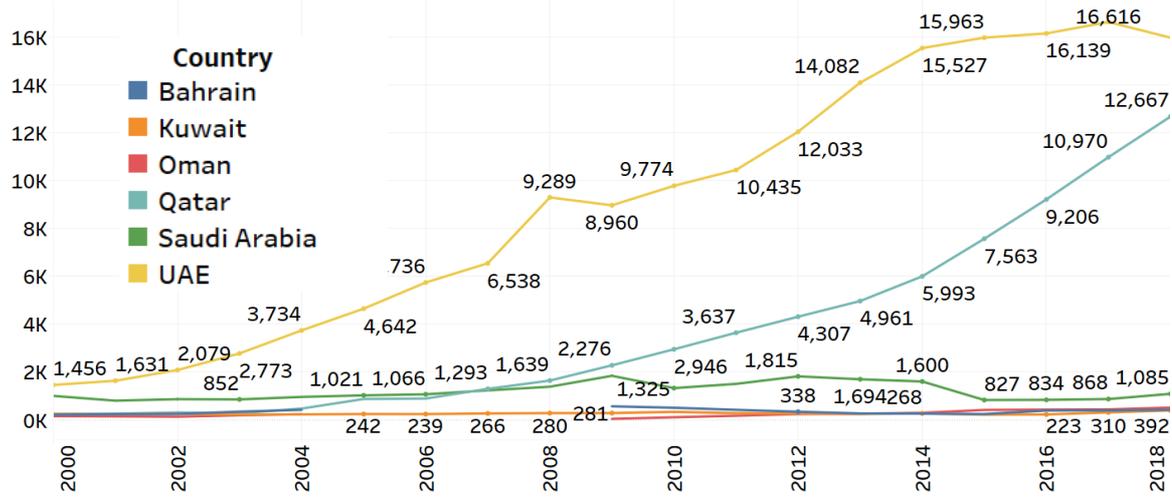

**Figure 2**. Air Transport, (passengers carried)

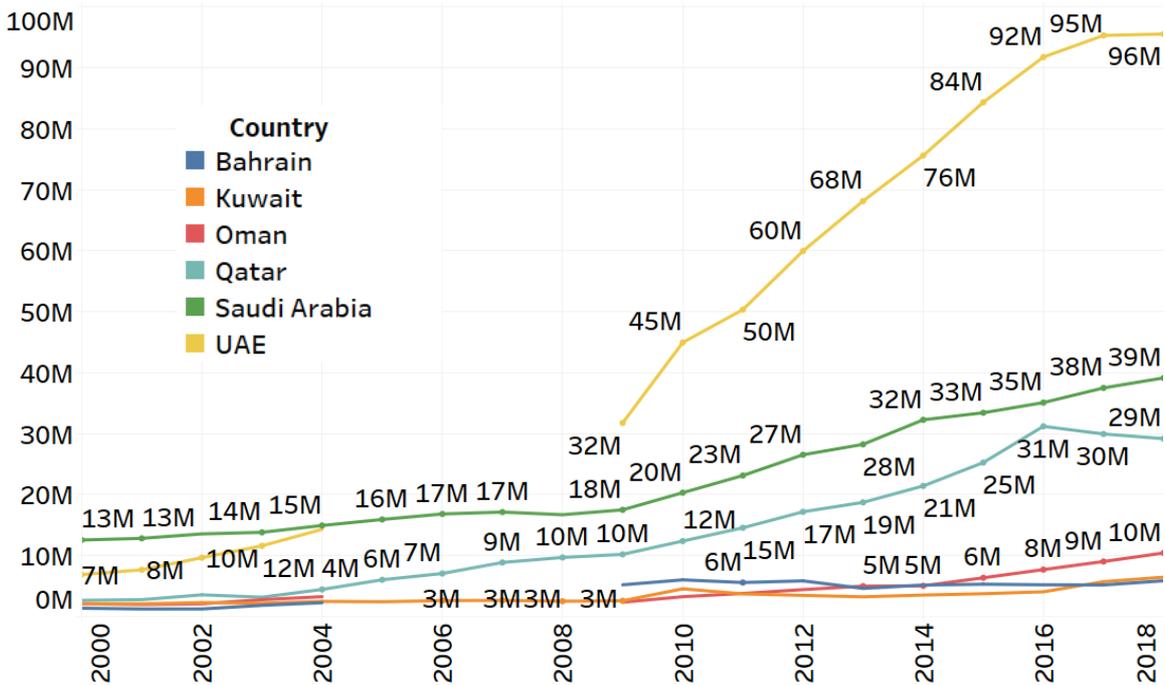

**Figure 3.** Manufacturing value added % of GDP

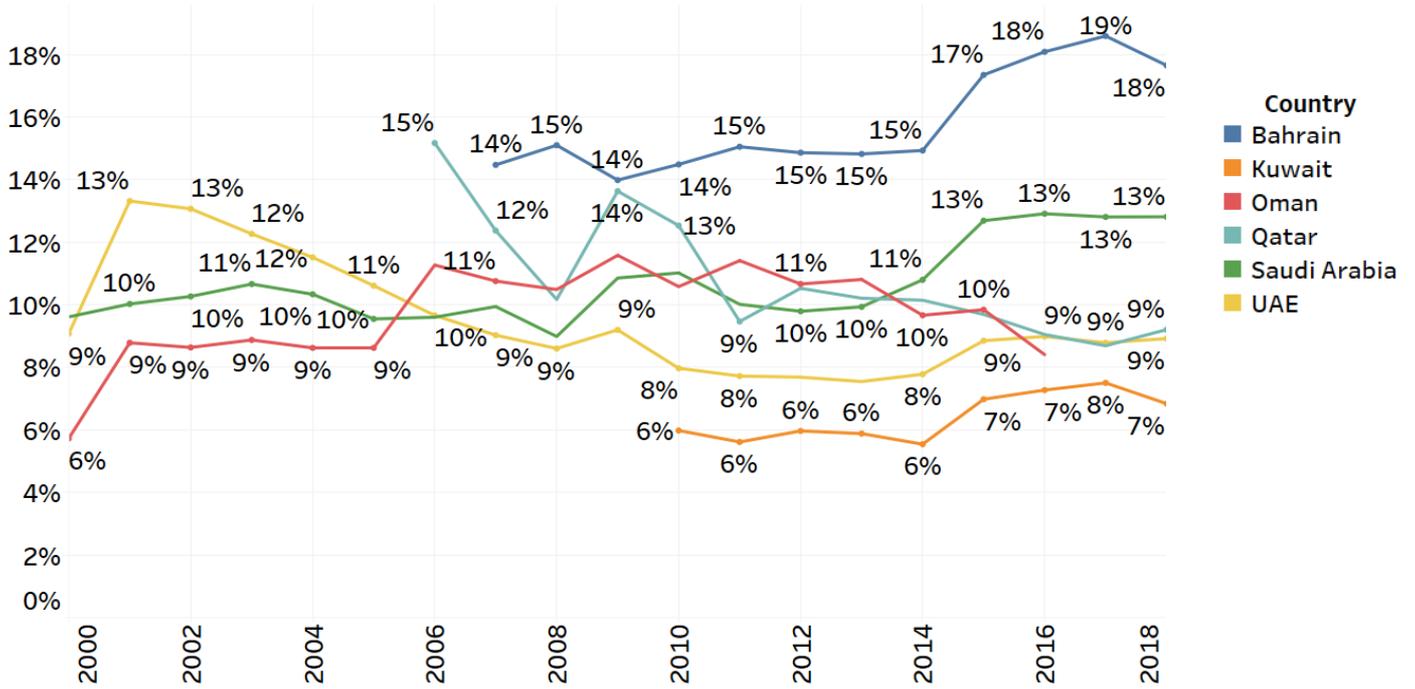

**Figure 4.** Employment in industry (% of total employment)

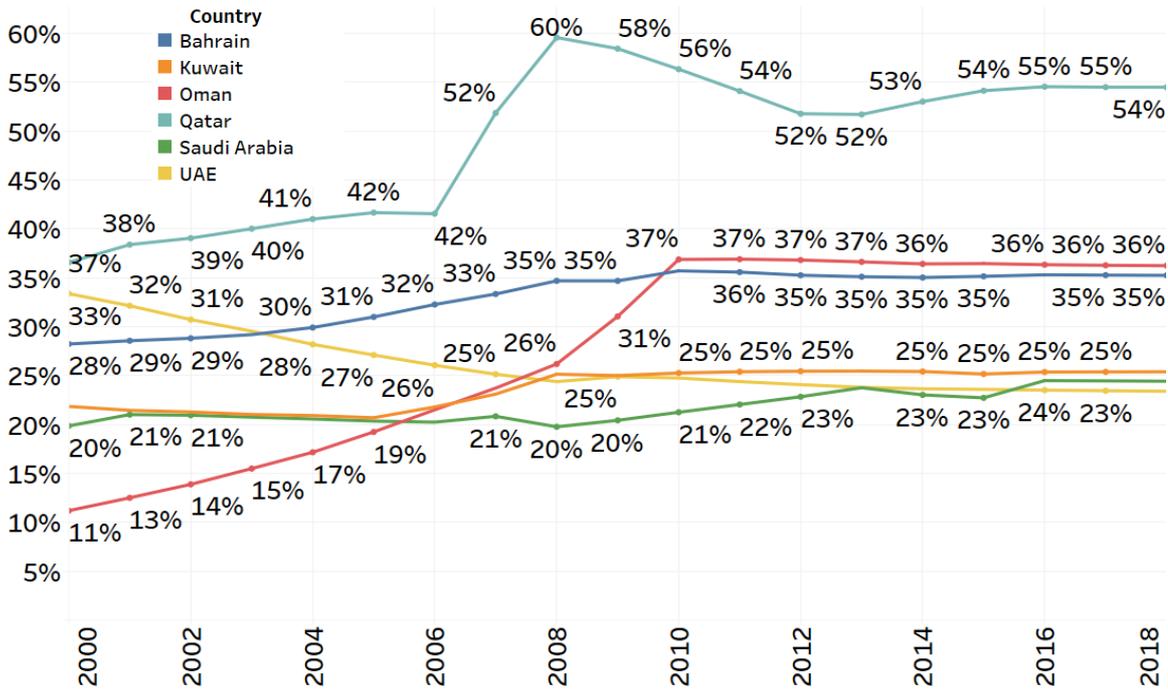

***Target 9.3: Increase access to financial services and markets.***

*SDG Indicator 9.3.1: Value of small-scale industry*

The UN data does not provide any indicators regarding the access of small-scale industry to financial services and markets. This indicator is defined as follows:

$$\text{Value of small} - \text{scale industry} = \frac{\text{small} - \text{scale industry}}{\text{Total industry}}$$

Achievement of the target is measured by measuring the share of small industries in the total value added in the industry. To this end, GCC countries have ambitious public incentives for innovative startups to include industrial companies with a more significant share in R&D programs.

*SDG Indicator 9.3.2: Small-scale industries with affordable credit*

The UN database does not provide any data on this indicator which is defined as follows:

$$\text{Value of small} - \text{scale industry with affordable credit}$$
$$= \frac{\text{small} - \text{scale industry with debt}}{\text{Value of small} - \text{scale industry}}$$

Progress in the target is measured by measuring the proportion of small industries that have a line of credit. Some GCC countries focused on improving bank guarantee procedures for small and medium-sized enterprises (SMEs) to obtain low-interest loans. For example, in the UAE, the Green Development Strategy emphasized the role of small and medium enterprises in developing the economy.

***Target 9.4: Upgrade all industries and infrastructures for sustainability.***

*SDG Indicator 9.4.1: CO2 emissions per unit value-added*

This indicator aims to measure the carbon intensity of the economies. Carbon dioxide ($CO_2$) intensity of economies is measured in kilograms of $CO_2$ per \$ of GDP (measured in international-\$ in 2011 prices). It is calculated as follows:

$$\text{CO2 emissions per unit value added}$$
$$= \frac{\text{Total carbon emissions in kilogram}}{GDP \text{ per capita measured in } 2011 \text{ } dollars}$$

This goal aims to motivate the businesses to upgrade infrastructure in order to make them more sustainable, with fewer carbon emissions. This will require increased resource-use efficiency. It will also promote companies to apply clean and environmentally sound technologies and industrial processes. However, in this part of the world, carbon is still at the core of the economic activities. It is the oil and gas reserves that brought the riches into GCC, and these sources continue to be the primary revenue source for the governments.

Figure 5 suggests that the carbon emission intensity of the Bahrain economy is the highest. The recent data on CO2 emissions per economic activity was 0.53 kilograms for each dollar of GDP. Bahrain does not have much oil. However, it runs a mega-scale refinery with Saudi Arabia, which is the reason for the high carbon intensity. Other countries in GCC are also carbon-intensive economies with ratios ranging from 0.35 to 0.39 kilograms of CO2 per each dollar of GDP. The persistence of carbon intensity is also another primary concern among these countries. The carbon footprint per economic activity around the world is declining, but this is not the case for GCC member states.

***Target 9.5: Enhance research and upgrade industrial technologies.***

*SDG Indicator 9.5.1: Research and Development (R&D) spending*

$$\text{Research and Development (R\&D) } spending$$
$$= \frac{\text{Spending on research and development}}{\text{GDP}}$$

The UN defines the expenditures on research and development as the current and capital expenditures on creative work undertaken to increase knowledge of the society. As such, it includes basic research, applied research, and experimental development. Progress in achieving the target is measured by R&D expenditure as a percentage of GDP. Spending on research and development as a share of GDP in UAE was 0.96% of GDP in 2016. Saudi Arabia has recently started allocating significant financial sources for R&D. The last available data for Saudi Arabia indicates 0.82% of its GDP was spend on R&D. Qatar is also catching up with them with a ratio of 0.5%. However, other GCC member states have very low R&D spending ratios, about 0.08% to 0.22%. Figure 6 shows that R&D spending is meager in GCC countries.

*SDG Indicator 9.5.2: Researchers per million inhabitants*

Progress in achieving this target is measured by the numbers of researchers located in individual states. UN defines researchers in Research & Development (R&D) as professionals engaged in the creation of new knowledge. Besides faculty members, postgraduate Ph.D. students are also included in the calculation of this indicator.

Figure 7 on the number of researchers show a wide range of differences between GCC members. The World Bank data suggests that Oman has only 244 researchers, and Kuwait has 492 researchers, whereas UAE hosts 2407 researchers per million people. However, the research output does not support those numbers, which renders them somewhat arbitrary numbers. Kuwait University alone hosts 1600 academic staff and 2200 postgraduate students. It is very likely that the variation in what is called a researcher is the reason for such difference. There is an increasing emphasis on boosting the number of researchers working in the government and private sectors. In the future, it is expected that the next generation of researchers will also be employed in research and development in manufacturing.

***Target 9.A: Facilitate sustainable infrastructure development for developing countries.***

*SDG Indicator 9.A.1: Development assistance for infrastructure*

While GCC members are among the large aid donors, the data is not

***Target 9.B: Support domestic technology development and industrial diversification.***

*SDG Indicator 9.B.1: Medium and high-tech industry*

Progress in achieving the target is measured by the percentage of value-added of the medium and advanced technological industry of total value-added. This is an indicator aimed at measuring whether the economy is moving towards high-tech industries or whether they are stalling with low tech, low productivity activities. The UN defines this indicator as the proportion of medium and high-tech industry (MHT) value-added as a percentage of total manufacturing value. Higher values indicate that a country's industrial sector is focused on medium to high-tech manufacturing. The values for GCC countries ranges between 20% (Oman) to 50% (Qatar). However, this indicator does not mean that GCC countries are high-tech countries. This indicator includes petrochemicals and construction industries in this category. Both of these medium-tech industrial categories are important for the region. That is why this indicator is quite high among GCC member states.

Figure 8 suggests that the GCC countries need to focus on capacity building in the industry by promoting foreign direct investments by multinational corporations as well as supporting SMEs focused on high-tech companies.

***Target 9.C: Access to information and communications technology.***

*SDG Indicator 9.C.1: Proportion of population covered by a mobile network*

This indicator measures the level of access to mobile communication networks, which includes cellular subscriptions as well as % of the population having access to the internet. The UN defines an individual who used the internet in the last three months as Internet users. The Internet can be used via any means. In GCC, internet access through mobile and optic-fiber subscriptions are fast and reliable. Figure 9 shows details on mobile subscriptions.

**Figure 5.** CO2 emissions (kg per 2011 dollar of GDP- PPP)

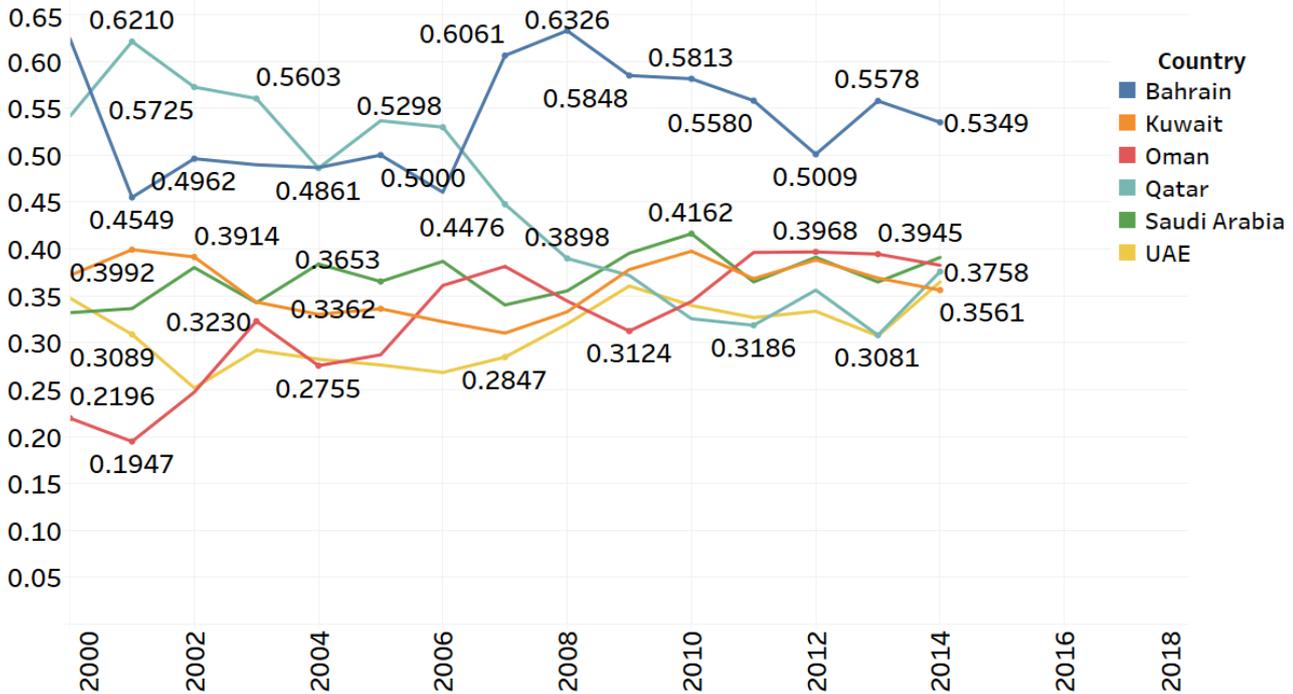

**Figure 6.** Research and Development Expenditure (% of GDP)

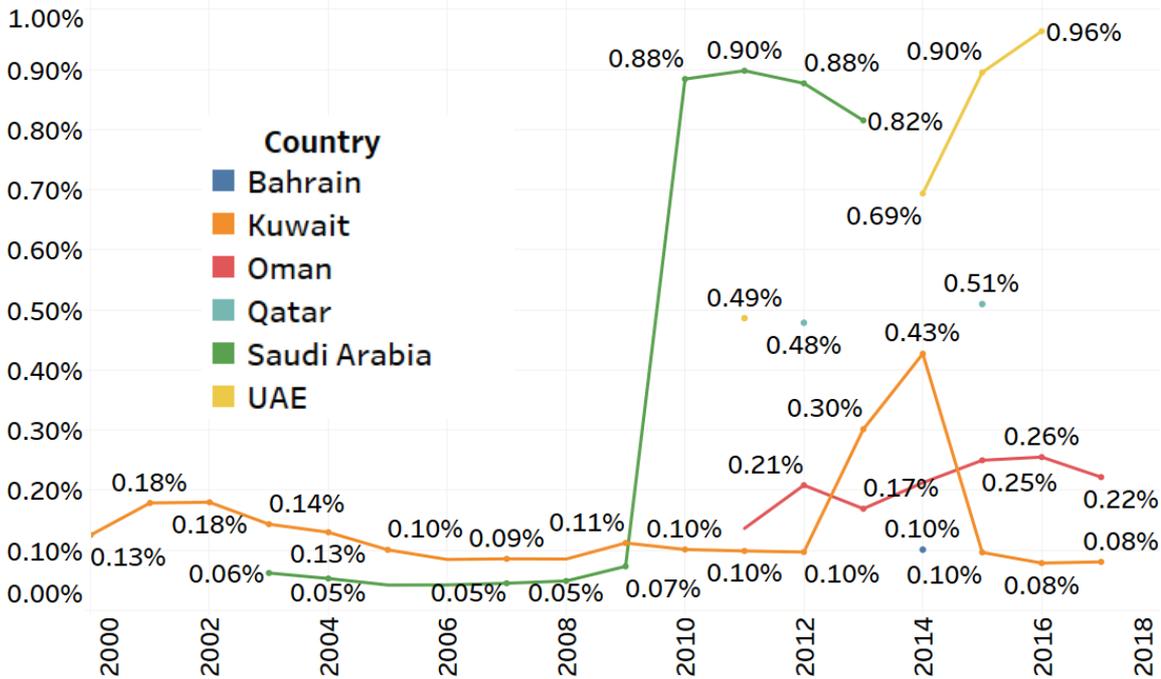

**Figure 7.** Researchers in R&D (per million people)

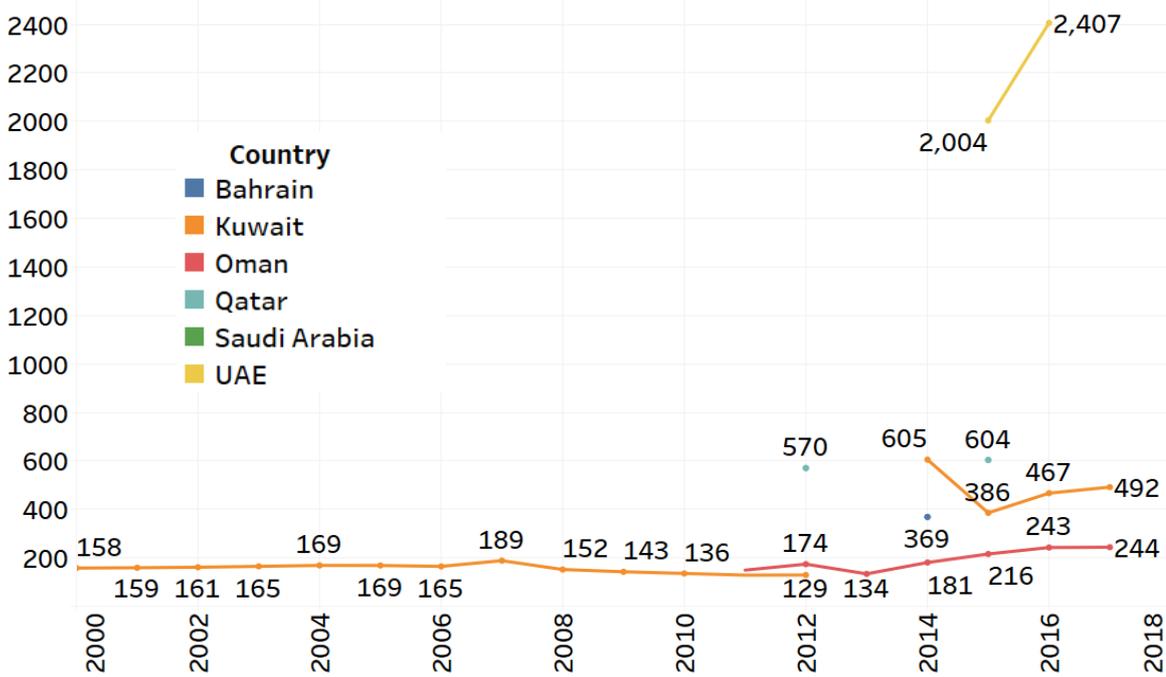

**Figure 8.** Medium and high-tech industry (% manufacturing value-added)

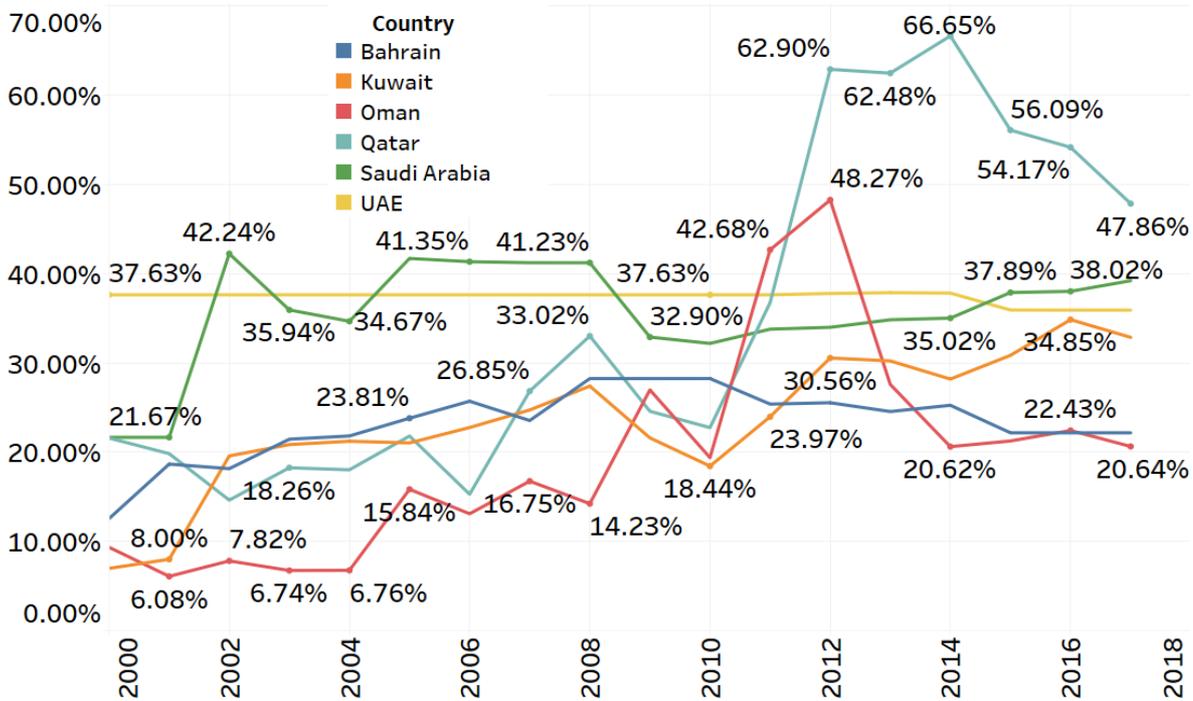

Many individuals use more than one mobile subscription for the service, whether that is prepaid or postpaid. The highest mobile cellular subscription is in UAE (208.5 per 100 people), whereas the lowest subscription rate is in Saudi Arabia (121.5 per 100 people). Internet usage is also quite high and reached almost 100% for all GCC members except Saudi Arabia and Oman. Figure 10 suggests that there is convergence to 100% internet access, and these countries also likely reach the 100% target soon.

**Analysis: Comparative GCC Performance on Achieving SDG 9**

While the UN declaration did not specify any methods to rank the countries, there is a significant interest in performing comparative analysis among countries. OECD and European commission have their ranking system for measuring progress towards SDGs. One ranking methodology that offers global coverage is created by Bertelsmann Stiftung (BS). This is a non-profit research organization headquartered in Germany. They introduced the BS-SDG index in 2015. Since then, BS-SDG has become an industry-standard method in measuring and comparing performance towards SDG attainments (Lafortune *et al.*, 2019). The indicators used in the BS-SDG index are chosen based on both their availability and robustness. There are also minor updates on indicators every year based on new research conducted. The latest report suggests six indicators for measuring performance in SDG 9, *industry, innovation, and infrastructure*. These indicators are briefly analyzed as follows:

*Population using the internet (% - sdg9_intuse)*

The UN defines this indicator as *the percentage of the total population who used the internet from any location in the last three months via a mobile network*. The source of this information is the World Development Indicators provided by the World Bank. BS defines this indicator in the same way as the UN.

*Mobile broadband subscriptions (per 100 inhabitants - sdg9_mobuse)*

The UN defines this indicator as *the percentage of the total population who used the internet from any location in the last three months via a mobile network*. Similar to the previous indicator, the source of this information is World Development Indicators provided by the World Bank. BS also defines this indicator in the same way as the UN.

*Logistics performance index (sdg9_lpi)*

This indicator is aimed at measuring the logistics infrastructure and is different from the UN indicators. While the UN measures the quality of logistics infrastructure in terms of the number of passenger numbers and freight volume, this indicator is based on the quality of trade and transport-related infrastructure. This is a survey-based assessment of the quality of trade and transport-related infrastructure, including information technology, on a scale from 1 (worst) to 5 (best).

*The Times Higher Education Universities Ranking (sdg_qs)*

This is another indicator aimed at measuring research performance and academic quality of higher education institutions in a country. While the UN measures the research and development in terms of research spending, this indicator directly measures the quality of the research conducted in the country. Known as THE rankings, this indicator evaluates the quality of the top 3 higher institutions located in the host country.

*The number of scientific and technical journal articles (per 1,000 population - sdg9_articles)*

This indicator is aimed at measuring the academic performance of the researcher in a country. Precisely, it is calculated as the number of scientific and technical journal articles published that are covered by the Science Citation Index (SCI) or the Social Sciences Citation Index (SSCI) based on the institutional address(es) listed in the article. The data is reported per capita.

*Research and development expenditure (% GDP - sdg9_rdex)*

The UN defines this indicator as to the *domestic expenditure on scientific research and experimental development (R&D) expressed as a percentage of Gross Domestic Product (GDP)*. The source of this information is also World Development Indicators provided by the World Bank. BS also defines this indicator in the same way as the UN.

*$CO_2$ emissions per unit of manufacturing value-added (sdg9_coemva)*

BS defines this indicator as *carbon dioxide emissions per unit of manufacturing value-added (kilograms of $CO_2$ per constant 2010 US$)*. This is almost in line with the UN definition. However, while the UN includes this in its annual report, BS does not include this data in its global coverage reports. It is only included in the recently initiated regional SDG report on Arab countries (Luomi *et al.*, 2019). However, some of the observations in this data do not make sense (such as Oman being the most carbon intense country); therefore, we decided not to report the numbers on this indicator.

The recent regional SDG report on Arab regions also suggests that while there are significant challenges in the GCC region, all six countries are improving their infrastructure and investing more resources towards research and development. The latest data on each of the above-listed indicators as defined by BS can be seen as in Table 2 below:

These numbers agree with the BS regional report on Arab states that GCC countries are among the higher performers in achieving SDG 9. We found that the GCC countries are on track or maintain their SDG achievements in SDG9. Communication and network infrastructure are robust. Each country has at least two telecommunication and high-speed internet service providers. One development strategy introduced in the UAE and followed by other GCC members is to establish free zones attached to mega-sized seaports and airports within the country. This strategy allowed the free zones to become commercial trading centers, thereby boosting the economies of the host countries.

A significant portion of the wealth gained by carbon-based natural capital has been invested in the education sector. Education is a highly profitable sector in GCC. Besides flagship national universities, GCC countries host several private colleges and universities. For example, there are more than 100 higher education institutions in the UAE alone. Most of these higher education institutes are connected with well-known international establishments in the Western world. Thus, universities play a significant role in boosting research and development within the GCC countries. However, in terms of publications, the academic output in GCC is well behind the global average. This is partially due to the lack of academic funding in this part of the world, which is also reflected in the low ratio of R&D expenditures as a percentage of GDP.

**Discussion and Policy Suggestions**

In this article, we analyzed the performance of GCC countries in achieving the SDG 9, *industry, innovation, and infrastructure*. Each SDG has subgoals to be measured by at least one indicator. The UN declaration for SDG 9 defined eight subgoals to be measured by 12 indicators. However, there was some ambiguity in the indicators defining some of the subgoals. Therefore, we also collect information from the BS-SDSN database. SDG 9 data on both the UN and BS-SDSN databases emphasis the role of industrial infrastructure, internet, and mobile connectivity, as well as the research and development activities. Based on the cross-comparison of the available data on SDG9, we can state the biggest strength of the GCC member states on achieving progress towards SDG 9 is logistics performance, internet, and mobile infrastructure.

**Figure 9**. Mobile Subscriptions (per 100 people)

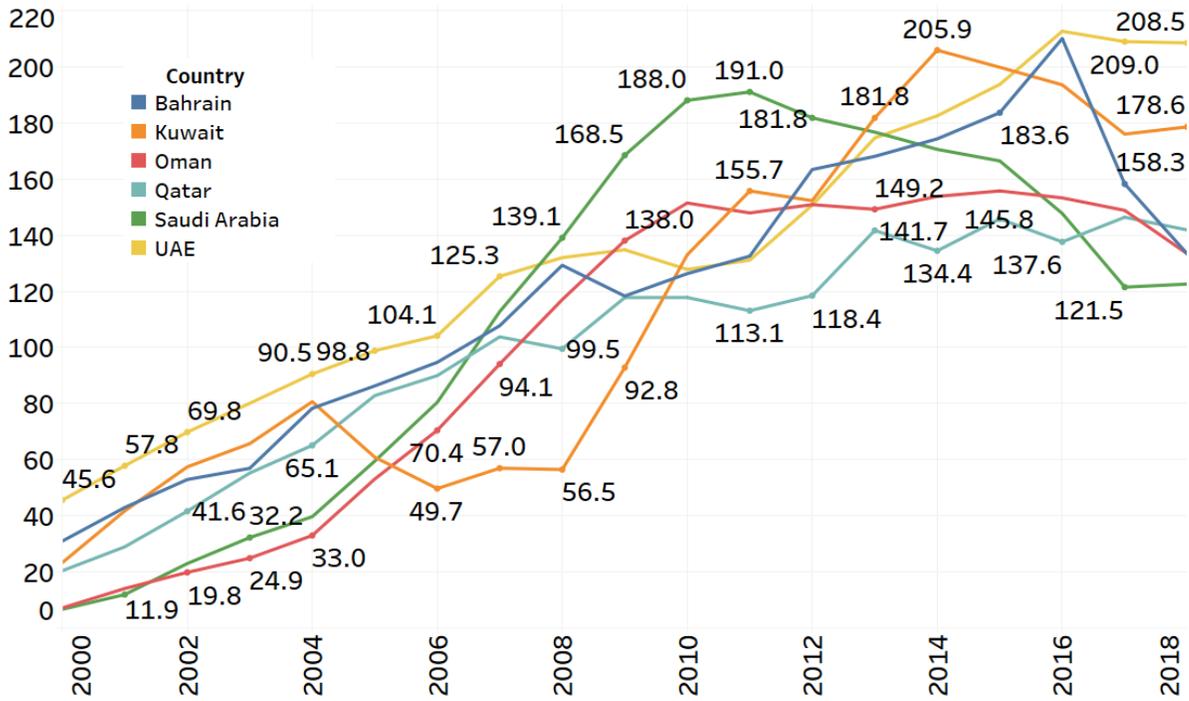

**Figure 10.** Individuals Using the Internet (% of Population)

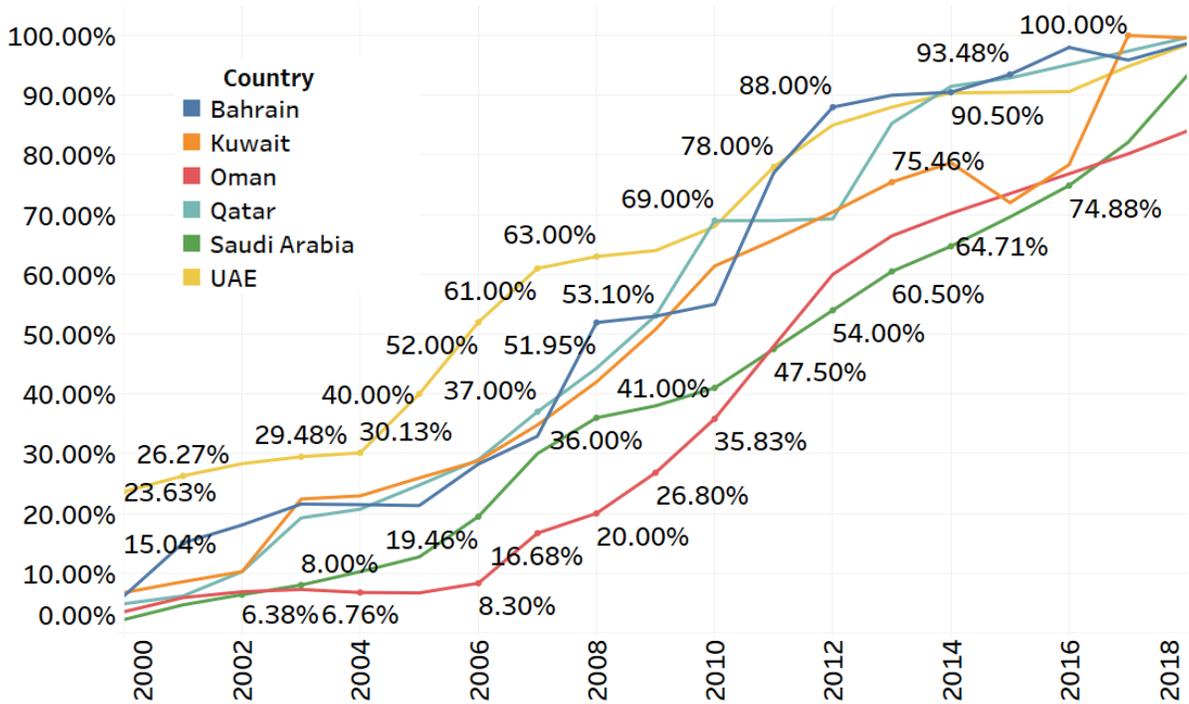

**Table 2.** BS-SDG Indicators

| Indicator | GCC Countries | | | | | | Global Values | | |
|---|---|---|---|---|---|---|---|---|---|
| | Bahrain | Kuwait | Oman | Qatar | Saudi Arabia | UAE | Mean | Min | Max |
| *sdg9_intuse* | 95.9% | 98% | 80.2% | 95.9% | 82.1% | 94.8% | 52.2% | 0 | 98.9% |
| *sdg9_mobuse* | 146 | 127.3 | 93.9 | 127.2 | 90 | 243.4 | 61% | 0 | 243% |
| *sdg9_lpi* | 2.7 | 3 | 3.2 | 3.4 | 3.1 | 4 | 2.7 | 1.6 | 4.4 |
| *sdg_qs* | 14.2 | 22.5 | 22.5 | 39.4 | 43.9 | 36.8 | 16.8 | 0 | 94.3 |
| *sdg9_articles* | 0.1 | 0.2 | 0.2 | 0.5 | 0.3 | 0.2 | 0.4 | 0 | 2.5 |
| *sdg9_rdex* | 0.1 | 0.4 | 0.2 | 0.5 | NA | 1 | 0.7 | 0 | 4.3 |

There is high competition among GCC member states to retrieve as much foreign investment as possible. There are also increasing emphasis to utilize the harsh weather conditions in favor of tourism, particularly during the winter months, when the weather is pleasant and allows for all kinds of outdoor activities. Each country hosts several free zones that are directly connected to either an airport or seaport. Moreover, each GCC country has at least one flagship air carrier supported by an associated airport on land.

- Bahrain has Gulf Air headquartered in Manama.
- Kuwait has Kuwait Airways headquartered in Kuwait City.
- Oman has Oman Air headquartered in Muscat.
- Qatar has Qatar Airways headquartered in Doha.
- Saudi Arabia has Saudia Airways headquartered in Jeddah.
- UAE has three major airlines: Emirates Airways headquartered in Dubai; Etihad Airways headquartered in Abu Dhabi; Arabian Airways headquartered in Sharjah.

There are also several charter and discount airways flying the region. While rail transport is still missing, GCC is well-connected with the rest of the world through its airports. Each GCC country also has at least one mobile and internet service provider offering high-speed internet access to the locals and expatriates alike.

On the negative side, while there is a significant investment in education, the research atmosphere is not in par with that of the Western world. It is true that there a few world-class publicly funded higher educational institutions providing a balance between research and teaching/service duties of the faculty members. However, from the private sector's perspective, education is seen as a customer-oriented business, where maximizing the return on investment is the primary motive. As there is a big competition to attract as many students as possible, student satisfaction ratings tend to shadow the research performance during the faculty evaluations, particularly in private higher educations. There is a need for state involvement in providing attractive and competitive research funding to support researchers in both private and public higher education institutions.

Denmark, the number ranked sustainability achiever sets a good role model for the GCC countries. Denmark is ranked as the number one achiever in SDG rankings, not only because it respects the environment but also its R&D expenditure as a percentage of GDP is 2.9%. This is between 5 to 30 times higher than the R&D expenditure ratios of GCC member states. Thanks to this investment in education, the average score of the top 3 universities in Denmark is 58.2, and the scientific output as measured by the number of articles is 2.4 (5 times higher than that of Qatar and 24 times higher than that of Bahrain). Denmark is also a founding member of the European Commission, which hosts the most extensive research program in the world. Known as the Horizon 2020, this research program has a budget of 80 billion Euros. Any researcher or group of researchers can apply to competitive research funding financed by the European Commission. The Commission promotes regional integration among academia by supporting student and faculty exchange within member states. Some specific research/training programs require at least 2-3 different institutions to apply together. In addition to the research program, the EU has committed 84 billion Euros for smart and inclusive growth, 60 billion Euros for sustainable growth, and natural resources (Council of the EU, 2019). A similar research scheme in the GCC area will not only boost research productivity, but it will also promote regional integration among the member states.

**Limitations and Future Research**

This article sets the foundation for conducting similar research relevant to other regional unions such as the European Union, African Union, CIS, and ASEAN member countries. However, our findings are highly dependent on the way that sustainability indicators are defined and measured. While international statistical agencies are trying to unify these indicators, they are almost exclusively dependent on the national statistical reports, which might be biased and inflated.

There is a significant motivation among all UN member countries to have a higher sustainability rank. These ranks are based on the sustainability indicators mentioned in this article. While the ultimate goal is to achieve sustainable development, the indicators measuring those goals might become the targets themselves. Therefore, for any comparative analysis, the data used shall be cross-checked with several other sources for confirmation.